\documentclass[pra,preprint]{revtex4}
\usepackage{graphicx}
\usepackage{amsfonts}
\usepackage{bm}

\begin{document}
\title{Practicable factorized  TDLDA for arbitrary density- and current-dependent
functionals}
\author{V.O. Nesterenko$^1$, J. Kvasil$^2$, and  P.-G. Reinhard$^3$}
\date{\today}
\address{$^{1}$
Bogoliubov Laboratory of Theoretical Physics,
Joint Institute for Nuclear Research, Dubna, Moscow region, 141980, Russia}
\address{$^{2}$
Department of Nuclear Physics, Charles University,
 CS-18000 Prague 8, Czech Republic}
\address{$^{3}$
Institut f\"ur Theoretische Physik II,
Universit\"at Erlangen, D-91058, Erlangen, Germany}

\begin{abstract}
We propose a practicable method for describing linear dynamics of different
finite Fermi systems. The method is based on a general self-consistent
procedure for factorization of the two-body residual interaction. It is
relevant for diverse density- and current-dependent functionals and, in fact,
represents the self-consistent separable random-phase approximation (RPA),
hence the name SRPA. SRPA allows to avoid diagonalization of high-rank RPA
matrices and thus dwarfs the calculation expense. Besides, SRPA expressions
have a transparent analytical form and so the method is very convenient
for the analysis and treatment of the obtained results. SRPA demonstrates high
numerical accuracy. It is very general and can be applied to diverse systems.
Two very different cases, the Kohn-Sham functional for atomic clusters and
Skyrme functional for atomic nuclei, are considered in detail as particular
examples. SRPA treats both time-even and time-odd dynamical variables and, in
this connection, we discuss the origin and properties of time-odd currents
and densities in initial functionals. Finally, SRPA is compared
with other self-consistent approaches for the excited states, including the
coupled-cluster method.
\end{abstract}

\maketitle

\arraycolsep1.5pt

\narrowtext



\section{Introduction}
\label{sec:introduction}

 The time-dependent local-density-approximation theory (TDLDA) is widely used for
description of dynamics of diverse quantum systems such as atomic nuclei, atoms
and molecules, atomic clusters, etc. (see for more details
\cite{Row_70,Ring_Schuck_80,Dreizler_90,Bertsch_Broglia_94, Rein_Suraud_03}).
However, even in the linear regime, this theory is plagued by dealing with
high-rank matrices which make the computational effort too expensive.
This is especially the case for non-spherical systems with their demanding
configuration space. For example, in the Random Phase Approximation (RPA),
a typical TDLDA theory for linear dynamics, the rank of the matrices is
determined by the size of the particle-hole 1ph space which becomes really huge
for deformed and heavy spherical systems. The simplest RPA versions, like
the sum rule approach and local RPA (see hierarchy of RPA methods in \cite{Rein_AP_92})
deal with a few collective variables instead of a full 1ph space and thus
avoid the problem of high-rank matrices. But these versions cannot properly describe
gross-structure of collective modes and the related property of the Landau damping
(dissipation of the collective motion over nearby 1ph excitations).

In this connection, we propose a method
\cite{Ne_PRA_98,Ne_AP_02,Miori_01,Ne_PRC_1d} which combines accuracy and power
of involved RPA versions with simplicity and physical transparency of the
simplest ones and thus is a good compromise between these two extremes. The
method is based on the self-consistent {\it separable} approximation for the
{\it two-body} residual interaction which is factorized  into a sum of weighted
products of {\it one-body} operators. Hence the method is called as separable
RPA (SRPA). It should be emphasized that the factorization is
self-consistent and thus does not result in any additional parameters.
Expressions for the one-body operators and their weights are inambiguously
derived from the initial functional. The factorization has the advantage to
shrink dramatically the rank of RPA matrix (usually from $r=10^3-10^6$ to
$r=2-14$) and thus to minimize the calculation expense. Rank of SRPA matrix is
determined by the number of the separable terms in the expansion for the
two-body interaction. Due to effective self-consistent procedure, usually a few
separable terms (or even one term) are enough for a good accuracy.
Ability of SRPA to minimize the computational effort becomes really decisive
in the case of non-spherical systems with its huge 1ph configuration space.
SRPA formalism is quite simple and physically transparent, which makes the method
very convenient for the analysis and treatment of the numerical results. Being
self-consistent, SRPA allows to extract spurious admixtures connected with
violation of the translational or rotational invariance. As is shown below, SRPA
exhibits accuracy of most involved RPA versions but for the much less expense.
Since SRPA exploits the full 1ph space, it equally well treats collective and
non-collective states and, what is very important, fully describes the Landau
damping, one of the most important properties of collective motion. SRPA is
quite general and can be applied to diverse finite Fermi systems (and thus to
different functionals), including those tackling both time-even densities and
time-odd currents. The latter is important not only for nuclear Skyrme
functionals \cite{Skyrme,Engel_75} which exploits a variety of time-even and
time-odd variables but also for electronic functionals whose generalized
versions deal with basic current densities (see e.g. \cite{KS_current}).

SRPA has been already applied for atomic nuclei and clusters, both spherical
and deformed. To study dynamics of valence electrons in atomic clusters, the
Konh-Sham functional \cite{KS,GL}was exploited
\cite{Ne_PRA_98,Ne_AP_02,Ne_EPJD_98,Ne_EPJD_02}, in some cases together with
pseudopotential and pseudo-Hamiltonian schemes \cite{Ne_EPJD_98}. Excellent
agreement with the experimental data \cite{HS_99} for the dipole plasmon
was obtained. Quite recently
SRPA was used to demonstrate a non-trivial interplay between Landau
fragmentation, deformation splitting and shape isomers in forming a profile of
the dipole plasmon in deformed clusters \cite{Ne_EPJD_02}.

In atomic nuclei, SRPA was derived \cite{Miori_01,Ne_PRC_1d,Prague_02} for the
demanding Skyrme functional involving a variety of densities and currents
(see \cite{Ben_RMP_03} for the recent review on Skyrme forces). SRPA
calculations for isoscalar and isovector giant resonances (nuclear counterparts
of electronic plasmons) in doubly magic nuclei demonstrated high accuracy of
the method \cite{Ne_PRC_1d}.

In the present paper, we give a detail, maybe even tutorial, description of SRPA,
consider and discuss its most important particular cases and compare it with
alternative approaches, including the equation-of-motion method in the coupled-cluster
theory. We thus pursue the aim to advocate SRPA for researchers from
other areas, e.g. from the quantum chemistry.

The paper is organized as follows. In Section \ref{sec:srpa}, derivation of
the the SRPA formalism is done. Relations of SRPA with other alternative
approaches are commented. In Sec. 3, the method to calculate SRPA strength
function (counterpart of the linear response theory) is outlined.
In Section \ref{sec:KS_Skyrme}, the
particular SRPA versions for the electronic Kohn-Sham  and nuclear Skyrme
functionals are specified and  the origin and role of time-odd
currents in functionals are scrutinized. In Sec. \ref{sec:choice}, the
practical SRPA realization is discussed. Some examples demonstrating
accuracy of the method in atomic clusters and nuclei are presented. The summary
is done in Sec. \ref{sec:summary}. In Appendix A, densities and currents for
Skyrme functional are listed. In Appendix B, the optimal ways to calculate
SRPA basic values  are discussed.

\section{Basic SRPA equations}
\label{sec:srpa}

RPA problem becomes much simpler if the residual two-body interaction is
factorized (reduced to a separable form)
\begin{equation}
\sum_{h_1,h_2,p_1,p_2} \!\!\! <h_2p_2|\hat{V}_{res}|p_1h_1> a^+_{p_1} a^+_{p_2} a_{h_2} a_{h_1}
\rightarrow
\sum_{k,k'=1}^{K} [
\kappa_{kk'} {\hat X}_k {\hat X}_{k'}+\eta_{kk'} {\hat Y}_k {\hat Y}_{k'}
]
\label{two_body}
\end{equation}
where
$$
{\hat X}_k = \sum_{ph} <p|{\hat X}_k |h> a^+_p a_h, \quad
{\hat Y}_k = \sum_{ph} <p|{\hat Y}_k |h> a^+_p a_h
$$
are time-even and time-odd one-body operators, respectively.
Further, $a^+_p$ ($a_h$) is the creation (annihilation) operator of the particle state $p$
(hole state $h$); $K$ is the number of the separable terms.

   Conceptually, the self-consistent procedure outlined below was first proposed
in \cite{LS_nuclei}.

\subsection{Time-dependent Hamiltonian}
\label{sec:tdh}

The system is assumed to undergo small-amplitude harmonic vibrations around
HF ground state. The starting point is a general time-dependent energy
functional
\begin{equation}
E(J_{\alpha}({\vec r},t))= \int {\cal H} (J_{\alpha}({\vec r},t)) d{\vec r}
\label{func_E}
\end{equation}
depending on an arbitrary set of densities and currents
defined through the corresponding operators as
\begin{equation}\label{J_alpha}
J_{\alpha}({\vec r},t)=
<\Psi(t) |\hat J_{\alpha}({\vec r})| \Psi(t)> =  \sum_{h}^{occ}
 \varphi^*_h({\vec r},t)\hat J_{\alpha}({\vec r})\varphi_h^{\mbox{}}({\vec r},t)
\label{J}
\end{equation}
where $\Psi(t)$ is the many-body function of the system as a Slater determinant,
and $\varphi^*_h$ is the wave function of the hole (occupied) single-particle state.
In general, the set (\ref{J}) includes both time-even and time-odd densities
and currents, see examples  in the Appendix A.

Time-dependent mean-field Hamiltonian directly follows from
(\ref{func_E})-(\ref{J_alpha}):
\begin{equation}
\hat{h} (\vec{r},t) \varphi_h
=
\frac{\delta {\cal H}}{\delta \varphi^*_h}
=
\sum_{\alpha} \frac{\delta {\cal H}}{\delta J_{\alpha}}
\frac{\delta J_{\alpha}}{\delta \varphi^*_h}
=
\sum_{\alpha} \frac{\delta {\cal H}}
{\delta J_{\alpha}} \hat{J}_{\alpha} \varphi_h .
\label{mfh}
\end{equation}

In the small-amplitude regime, the densities are decomposed into static part and small
time-dependent variation
\begin{equation}
J_{\alpha}({\vec r},t) = \bar{J}_{\alpha}({\vec r}) + \delta J_{\alpha}({\vec r},t) .
\label{dJ}
\end{equation}
Then, to the linear order for $\delta J_{\alpha}({\vec r},t)$,
the mean-field Hamiltonian (\ref{mfh}) can be
decomposed into static and time-dependent response parts
\begin{eqnarray}
\hat{h}(\vec r,t) &=&  \hat{h}_0(\vec r) + \hat{h}_{res}(\vec r,t), \nonumber \\
                  &=&  \sum_{\alpha}
[\frac{\delta {\cal H}}{\delta J_{\alpha}}]_{J=\bar{J}}
{\hat J}_{\alpha}({\vec r})
+ \sum_{\alpha, \alpha'}
[\frac{\delta^2 {\cal H}}
{\delta J_{\alpha}\delta J_{\alpha'}}]_{J=\bar{J}}
\delta J_{\alpha'}({\vec r},t)
{\hat J}_{\alpha}({\vec r})
\label{eq:h_resp}
\end{eqnarray}
and thus we get the time-dependent Hamiltonian $\hat{h}_{res}(\vec r,t)$ responsible for the
collective motion.

\subsection{Scaling perturbation}
\label{sec:scaling_wf}

Now we should specify the response Hamiltonian
$\hat{h}_{res}(\vec r,t)$. For this aim, we use the scaling transformation
and define the perturbed many-body wave function of the system as
\begin{equation}\label{eq:scaling}
|\Psi(t)\!>=\prod_{k=1}^K
exp[-iq_{k}(t)\hat{P}_{k}]
exp[-ip_{k}(t)\hat{Q}_{k}]
|0> .
\end{equation}
Here both the perturbed wave function $|\Psi(t)\!>$ and static ground state
wave function $|0>$ are Slater determinants; ${\hat Q}_{k}(\vec{r})$
and ${\hat P}_{k}(\vec{r})$ are generalized coordinate (time-even) and momentum
(time-odd) hermitian operators with the propoerties.
\begin{eqnarray}\label{eq:P_Q}
\hat{Q}_{k} =\hat{Q}_{k}^+,\quad
\hat{T}\hat{Q}_{k}\hat{T}^{-1}&=&\hat{Q}_{k}, \nonumber \\
\hat{P}_{k} = i[\hat{H},\hat{Q}_{k}]_{ph}=\hat{P}_{k}^+,\quad
\hat{T}\hat{P}_{k}\hat{T}^{-1}&=&-\hat{P}_{k},\quad
\end{eqnarray}
They generate T-even and T-odd harmonic deformations $q_{k}(t)$ and
$p_{k}(t)$; $\hat{T}$ is the time inversion operator.

Using Eqs. (\ref{J}) and (\ref{eq:scaling}),
the transition densities read
\begin{eqnarray}\label{eq:trans_dens}
\delta J_{\alpha}({\vec r},t)&&  =
<\Psi(t)|{\hat J}_{\alpha}|\Psi(t)> -
<0|{\hat J}_{\alpha}|0>  =  \\
&&  =
i \sum_k
\{ q_{k}(t)<0|[\hat{P}_{k},\hat J_{\alpha}(\vec r)]|0>
+ p_{k}(t)<0|[\hat{Q}_{k},\hat J_{\alpha}(\vec r)]|0>\}
\nonumber
\end{eqnarray}
and the response Hamiltonian (\ref{eq:h_resp}) is
\begin{equation}\label{eq:h_XY}
\hat{h}(\vec r,t)  = \sum_{s k}
\{ q_{k}(t) \hat{X}_{k}({\vec r})+
 p_{k}(t) \hat{Y}_{k}({\vec r})\}
\end{equation}
where all the ${\vec{r}}$-dependent terms are collected
into the hermitian one-body operators
\begin{eqnarray}\label{eq:X}
\hat{X}_{k}({\vec r})   &=&
i\sum_{\alpha \alpha'}
[\frac{\delta^2 {\cal H}} {\delta J_{\alpha}
\delta J_{\alpha'}}]_{J=\bar{J}}
<0|[\hat{P}_{k} ,{\hat J}_{\alpha'}]|0>
{\hat J}_{\alpha}(\vec r), \\
\hat{Y}_{k}({\vec r})   &=&
i\sum_{\alpha \alpha'}
[\frac{\delta^2 {\cal H}} {\delta J_{\alpha}
\delta J_{\alpha'}}]_{J=\bar{J}}
<0|[\hat{Q}_{k} ,{\hat J}_{\alpha'}]|0>
{\hat J}_{\alpha}(\vec r)
\label{eq:Y}
\end{eqnarray}
with the properties
\begin{eqnarray}
\hat{X}_{k} &=& \hat{X}_{k}^+,  \quad
T\hat{X}_{k}T^{-1} = \hat{X}_{k}, \quad
\hat{X}_k^* = \hat{X}_k,
\\
\hat{Y}_{k} &=& \hat{Y}_{k}^+, \quad
T\hat{Y}_{k}T^{-1} = -\hat{Y}_{k}, \quad
\hat{Y}_k^*=-\hat{Y}_k .
\end{eqnarray}
As is shown below, ${\hat X}_k$  and ${\hat Y}_k$ are just the T-even and
T-odd operators to be exploited in the separable expansion (\ref{two_body}).

In the derivation above, we used the property of hermitian operators
with a definite T-parity
\begin{equation}\label{eq:<[A,B]>=0}
  <0|[\hat{A},\hat{B}]|0> = 0, \quad \mbox{if} \quad
T\hat{A}T^{-1} = T\hat{B}T^{-1}
\end{equation}
which states that the average of the commutator vanishes if the operators
$\hat{A}$ and $\hat{B}$ are of the same T-parity. This property allows to
classify operators with a definite T-parity in the SRPA formalism
and thus to make the formalism simple and transparent. For example, in Eqs.
(\ref{eq:X})-(\ref{eq:Y}), T-even and T-odd densities ${\hat J}_{\alpha}(\vec r)$
contribute separately to ${\hat X}_k$ and ${\hat Y}_k$.

To complete the construction of the separable expansion (\ref{two_body}),
we should yet determine the strength matrices
$\kappa_{kk'}$ and  $\eta_{kk'}$. This can be done through
variations of the basic operators
\begin{eqnarray}
\delta {\hat X}_{k}(t)  && \equiv
<\Psi(t)|{\hat X}_{k}|\Psi(t)>-<0|{\hat X}_{k}|0> =
\\
&& = i \sum_{k'}
q_{k'}(t)<0|[\hat{P}_{k'},\hat{X}_{k}]|0> =
- \sum_{k'} q_{k'}(t) \kappa_{k'k}^{-1} \; ,
\nonumber
\label{eq:Xvar}
\end{eqnarray}
\begin{eqnarray}
\delta {\hat Y}_{k}(t)  && \equiv
<\Psi(t)|{\hat Y}_{k}|\Psi(t)>-<0|{\hat Y}_{k}|0> =
\\
&& = i \sum_{k'} p_{k'}(t) <0|[\hat{Q}_{k'},\hat{Y}_{k}]|0> =
- \sum_{k'} p_{k'}(t) \eta_{k'k}^{-1}
\nonumber
\label{eq:Yvar}
\end{eqnarray}
where we introduce symmetric inverse strength matrices
\begin{eqnarray}
\label{eq:kappa}
\kappa_{k'k}^{-1 }&&=
\kappa_{kk'}^{-1} =
- i <0|[\hat{P}_{k'},{\hat X}_{k}]|0> = \\
&& = \int d{\vec r}\sum_{\alpha \alpha'}
[\frac{\delta^2 {\cal H}}{\delta J_{\alpha'}\delta J_{\alpha}}]
<0|[\hat{P}_{k},{\hat J}_{\alpha}]|0>
<0|[\hat{P}_{k'},{\hat J}_{\alpha'}]|0>
,\nonumber
\end{eqnarray}
\begin{eqnarray}
\label{eq:eta}
\eta_{k'k}^{-1 }&&=
\eta_{kk'}^{-1 } = -i
<0|[\hat{Q}_{k'},{\hat Y}_{k}]|0>  \\
&& = \int d{\vec r}\sum_{\alpha \alpha'}
[\frac{\delta^2 {\cal H}} {\delta J_{\alpha'}\delta J_{\alpha}}]
<0|[\hat{Q}_{k},{\hat J}_{\alpha}]|0>
<0|[\hat{Q}_{k'},{\hat J}_{\alpha'}]|0>.  \nonumber
\end{eqnarray}
Then one gets
\begin{eqnarray}
- \sum_{k} \kappa_{k'k} \delta {\hat X}_{k}(t) &=& q_{k'}(t) ,
\\
-  \sum_{sk} \eta_{k'k} \delta {\hat Y}_{k}(t) &=& p_{k'}(t)
\end{eqnarray}
and the response Hamiltonian (\ref{eq:h_XY})
acquires  the form
\begin{equation}\label{eq:h_dXdY}
\hat{h}(\vec r,t)  =  - \sum_{kk'}
\{
\kappa_{kk'} \delta\hat{X}_{k}(t) \hat{X}_{k'}({\vec r})+
\eta_{kk'} \delta\hat{Y}_{k}(t) \hat{Y}_{k'}({\vec r})
\} .
\end{equation}
Following \cite{Row_70}, the response Hamiltonian (\ref{eq:h_dXdY}) leads
to the same eigenvalue problem as the separable Hamiltonian
\begin{equation}
{\hat H}_{RPA}={\hat h}_0 + {\hat V}_{res},
\label{eq:H_sep}
\end{equation}
with
\begin{eqnarray}
{\hat V}_{res}= -\frac {1}{2} \sum_{kk'}
[\kappa_{kk'} {\hat X}_{k} {\hat X}_{k'}
+ \eta_{kk'} {\hat Y}_{k} {\hat Y}_{k'}]
\label{eq:V_res_sep}
\end{eqnarray}
(see also discussion in the next subsections).

In principle, we already have in our disposal the SRPA formalism for
description of the collective motion
in space of collective variables. Indeed, Eqs. (\ref{eq:X}), (\ref{eq:Y}),
(\ref{eq:kappa}), and (\ref{eq:eta})
deliver one-body operators and strength matrices we need
for the separable expansion of the two-body interaction.
The number K of the collective variables $q_{k}(t)$ and $p_{k}(t)$
and separable terms depends on how precisely we want to describe
the collecive motion (see discussion in Section 4).
For $K=1$, SRPA converges to the sum rule approach
with a one collective mode  \cite{Rein_AP_92}.
For $K>1$, we have a system of K coupled oscillators and SRPA
is reduced to the local RPA \cite{Re_PRA_90,Rein_AP_92} suitable for a
rough description of several modes and or main gross-structure efects.
However, SRPA is still not ready to describe the Landau fragmentation. For
this aim, we should consider the detailed 1ph space. This will be done in the next
subsection.

\subsection{Introduction of 1ph space}
\label{sec:1ph}

Collective modes can be viewed as superpositions of 1ph configurations.
It is convenient to define this relation by using the Thouless theorem which
establishes the connection between two arbitrary Slater determinants
\cite{Thouless}. Then, the perturbed many-body wave function reads
\begin{equation}
|\Psi(t)> = (1+\sum_{ph} c_{ph}(t) \hat{A}^+_{ph}) |\Psi_0>
\label{Psi_Thouless}
\end{equation}
where
\begin{equation}
 \hat{A}^+_{ph}=a^{\dagger}_p a_h
\label{A}
\end{equation}
is the creation operator of 1ph pair and
\begin{equation}
c_{ph}(t)=c^{+}_{ph}e^{i\omega t}+c^{-}_{ph}e^{-i\omega t}
\label{eq:c_ph}
\end{equation}
is the harmonic time-dependent particle-hole amplitude.
T-even $q_{k}(t)$ and T-odd $p_{k}(t)$ collective variables can be
also specified  as harmonic oscillations
\begin{eqnarray}
q_{k}(t)=\bar{q}_{k} cos(\omega t)
=\frac{1}{2}\bar{q}_{k}(e^{i\omega t}+e^{-i\omega t}), \nonumber \\
p_{k}(t)=\bar{p}_{k} sin(\omega t)=
\frac{1}{2i}\bar{p}_{k}(e^{i\omega t}-e^{-i\omega t}).
\label{eq:qp_harm}
\end{eqnarray}

Substituting  (\ref{eq:h_XY})  and (\ref{Psi_Thouless})
into the time-dependent HF equation
\begin{equation}
i\frac{d}{dt}|\Psi(t)>=
({\hat h}_0+{\hat h}_{res}(t))|\Psi(t)> ,
\label{eq:HF}
\end{equation}
one gets, in the linear approximation, the relation between
$c^{\pm}_{ph}$ and collective deformations $\bar{q}_{k}$ and $\bar{p}_{k}$
\begin{equation}
c^{\pm}_{ph}=-\frac{1}{2}
\frac{\sum_{k'} [\bar{q}_{k'}
<ph|\hat{X}_{k'}|0>
\mp i \bar{p}_{k'}<ph|\hat{Y}_{k'}|0>]}
{\varepsilon_{ph}\pm\omega},
\label{eq:c_pm_qp}
\end{equation}
where $\varepsilon_{ph}$ is the energy of 1ph pair.

In addition to Eqs. (\ref{eq:Xvar})-(\ref{eq:Yvar}), the variations
$\delta {\hat X}_{k}(t)$  and $\delta {\hat Y}_{k}(t)$
can be now  obtained with the alternative perturbed wave function
(\ref{Psi_Thouless}):
\begin{eqnarray}
\delta {\hat X}_{k'}(t) && \!\!\!\!\! =  \!\!
\sum_{ph}
(c_{ph}(t)^*<\!ph|\hat{X}_{k'}|0\!>
+c_{ph}(t)<\!0|\hat{X}_{k'}|ph\!>),
\label{eq:dX_ph}  \\
\delta {\hat Y}_{k'}(t) && \!\!\!\!\!  =  \!\!
\sum_{ph}
(c_{ph}(t)^*<\!ph|\hat{Y}_{k'}|0\!>+
c^s_{ph}(t)<\!0|\hat{Y}_{k'}|ph\!>).
\label{eq:dY_ph}
\end{eqnarray}
It is natural to equate  the dynamical variations of the basic operators
$\delta{\hat X}_{k}$ and $\delta{\hat Y}_{k}$, obtained with the scaling
(\ref{eq:scaling}) and Thouless (\ref{Psi_Thouless})  perturbed wave functions.
This provides the additional relation between the amplitudes
$c^{\pm}_{ph}$ and deformations $\bar{q}_{k}$ and  $\bar{p}_{k}$
and finally result in the system of equations for the unknowns
$\bar{q}_{k}$ and  $\bar{p}_{k}$.

By equating  (\ref{eq:Xvar})-(\ref{eq:Yvar})
and (\ref{eq:dX_ph})-(\ref{eq:dY_ph})  we get
\begin{eqnarray}
\label{eq:X_c_qp}
- \sum_{k} q_{k}(t) \kappa_{kk'}^{-1}
&=&
\sum_{ph}
(c_{ph}(t)^*<\!ph|\hat{X}_{k'}|0\!>
+c_{ph}(t)<\!0|\hat{X}_{k'}|ph\!>),
\\
- \sum_{k} q_{k}(t) \eta_{kk'}^{-1}
&=&
 \sum_{ph}
(c_{ph}(t)^*<\!ph|\hat{Y}_{k'}|0\!>+
c_{ph}(t)<\!0|\hat{Y}_{k'}|ph\!>).
\label{eq:Y_c_qp}
\end{eqnarray}
Substituting  (\ref{eq:c_ph})-(\ref{eq:c_pm_qp})
into these expressions and collecting, for example, the terms at
$e^{i\omega t}$, one finally gets
 \begin{eqnarray}
\label{eq:RPA_eq}
\sum_{k}  &
\{
\bar{q}_{k}
[F_{k'k}^{(XX)}- \kappa_{kk'}^{-1}]
+\bar{p}_{k} F_{k'k}^{(XY)}
\}=0 ,
\nonumber\\
\sum_{k}  &
\{
\bar{q}_{k}
F_{k'k}^{(YX)}
+\bar{p}_{k}
[F_{k'k}^{(YY)} - \eta_{kk'}^{-1}]
\}=0
\end{eqnarray}
with
\begin{eqnarray}
\label{eq:F_XX}
F_{k'k}^{(XX)} && =
 \sum_{ph}
\frac{1}{\varepsilon_{ph}^2-\omega^2}
\{<ph|\hat{X}_{k}|0>^* <ph|\hat{X}_{k'}|0>
(\varepsilon_{ph}+\omega)  \\
&& \hspace{3.5cm}+
<ph|\hat{X}_{k}|0> <0|\hat{X}_{k'}|ph>
(\varepsilon_{ph}-\omega)\} ,
\nonumber \\
\label{eq:F_YX}
F_{k'k}^{(YX)} && =
-i \sum_{ph}
\frac{1}{\varepsilon_{ph}^2-\omega^2}
\{<ph|\hat{Y}_{k}|0>^* <ph|\hat{X}_{k'}|0>
(\varepsilon_{ph}+\omega)\\
&& \hspace{3.5cm}+
<ph|\hat{Y}_{k}|0> <0|\hat{X}_{k'}|ph>
(\varepsilon_{ph}-\omega)\} ,
\nonumber \\
\label{eq:F_XY}
F_{k'k}^{(XY)} && =
i \sum_{ph}
\frac{1}{\varepsilon_{ph}^2-\omega^2}
 \{<ph|\hat{X}_{k}|0>^* <ph|\hat{Y}_{k'}|0>
(\varepsilon_{ph}+\omega) \\
&& \hspace{3.5cm}  +
<ph|\hat{X}_{k}|h> <0|\hat{Y}_{k'}|ph>
(\varepsilon_{ph}-\omega)\} ,
\nonumber\\
\label{eq:F_YY}
F_{k'k}^{(YY)} && =
  \sum_{ph}
\frac{1}{\varepsilon_{ph}^2-\omega^2}
\{<ph|\hat{Y}_{k}|0>^* <ph|\hat{Y}_{k'}|0>
(\varepsilon_{ph}+\omega)
 \\
&& \hspace{3.5cm}+
<ph|\hat{Y}_{k}|0> <0|\hat{Y}_{k'}|ph>
(\varepsilon_{ph}-\omega)\} .
\nonumber
\end{eqnarray}
Equating determinant of the system (\ref{eq:RPA_eq}) to zero,
we get the dispersion equation for RPA eigenvalues $\omega_{\nu}$.

\subsection{Normalization condition}

By definition, RPA operators of excited one-phonon states read
\begin{equation}\label{eq:Q_op}
\hat{Q}^+_{\nu}=\frac{1}{2}\sum_{ph}
\{
c^{\nu -}_{ph} \hat{A}^+_{ph} - c^{\nu +}_{ph} \hat{A}_{ph}
\}
\end{equation}
and fulfill
\begin{equation}\label{eq:Q_norm}
[\hat{Q}_{\nu},\hat{Q}_{\nu'}^+] = \delta_{\nu,\nu'}, \quad
[\hat{Q}_{\nu}^+,\hat{Q}_{\nu'}^+] = [\hat{Q}_{\nu},\hat{Q}_{\nu'}] = 0 ,
\end{equation}
where $\hat{A}^+_{ph}$ and $c^{\nu \pm}_{ph}$
are given by (\ref{A}) and (\ref{eq:c_pm_qp}), respectively.
In the quasiboson approximation for $\hat{A}^+_{ph}$, the normalization condition
$[\hat{Q}_{\nu},\hat{Q}_{\nu}^+] = 1$ results in the relation
\begin{equation}\label{eq:c_norm}
\sum_{ph}
\{
(c^{\nu -}_{ph})^2 - (c^{\nu +}_{ph})^2
\} =2 .
\end{equation}
Using (\ref{eq:c_pm_qp}),  it can be reformulated in terms
of the RPA matrix coefficients (\ref{eq:F_XX})-(\ref{eq:F_YY}):
\begin{eqnarray}\label{eq:c_norm_F}
&&  \sum_{ph}
\{
(c^{\nu -}_{ph})^2 - (c^{\nu +}_{ph})^2
\}
 \\
&& =  \sum_{kk'}\frac{1}{4}
 \{
\bar{q}^{\nu}_{k'} \bar{q}^{\nu}_{\bar{k}}
\frac{\partial F_{k'k}^{(XX)}(\omega_{\nu})}{\partial\omega_{\nu}}
+
2 \bar{q}^{\nu}_{k'} \bar{p}^{\nu}_{\bar{k}}
\frac{\partial F_{k'k}^{(YX)}(\omega_{\nu})}{\partial\omega_{\nu}}
+
\bar{p}^{\nu}_{k'} \bar{p}^{\nu}_{\bar{k}}
\frac{\partial F_{k'k}^{(YY)}(\omega_{\nu})}{\partial\omega_{\nu}}
\} = 2 N_{\nu} .
\nonumber
\end{eqnarray}
The variables  $\bar{q}^{\nu}_{k}$ and
$\bar{p}^{\nu}_{k}$ should be finally normalized
by the factor $1/\sqrt{N_{\nu}}$.

\subsection{General discussion}

Eqs. (\ref{eq:X}), (\ref{eq:Y}), (\ref{eq:kappa}), (\ref{eq:eta}),
(\ref{eq:c_pm_qp}), (\ref{eq:RPA_eq})-(\ref{eq:F_YY}), and
(\ref{eq:Q_op})-(\ref{eq:c_norm_F}) constitute the basic SRPA formalism.
It is worth now to comment some essential points:
\\
$\bullet$
One may show (e.g. by using a standard derivation of the matrix RPA)
that the separable Hamiltonian (\ref{eq:H_sep})-(\ref{eq:V_res_sep})
with (\ref{eq:Q_op}) results in the SRPA equations (\ref{eq:RPA_eq})-(\ref{eq:F_YY})
if to express unknowns $c^{\nu \pm}_{ph}$ through $\bar{q}_{\bar{k}}$ and
$\bar{p}_{\bar{k}}$.
Generally, RPA equations for unknowns $c^{\nu \pm}_{ph}$
require the RPA matrix of the high rank equal to size of the
$1ph$ basis. The separable approximation allows to reformulate the
RPA problem in terms of much more compact unknows $\bar{q}_{\bar{k}}$ and
$\bar{p}_{\bar{k}}$ (see relation (\ref{eq:c_pm_qp})
and thus to minimize the computational effort. As is seen from (\ref{eq:RPA_eq}),
the rank of the SRPA matrix is equal to a double number $K$ of the separable
operators and hence is low.
\\
$\bullet$
The number of RPA eigen-states $\nu$ is equal to the number of the relevant $1ph$
configurations used in the calculations. In heavy nuclei and atomic clusters, this
number ranges the interval $10^3$-$10^6$.
For every RPA state $\nu$, Eq. (\ref{eq:RPA_eq}) delivers a particular set of the
amplitudes $\bar{q}^{\nu}_{sk}$ and $\bar{p}^{\nu}_{sk}$ which,
following Eq. (\ref{eq:c_pm_qp}), self-consistently
regulate relative contrubutions of different
T-even and T-odd oscillating densities to the $\nu$-state.
\\
$\bullet$
Eqs. (\ref{eq:X}), (\ref{eq:Y}), (\ref{eq:kappa}), (\ref{eq:eta}) relate
the basic SRPA values with the starting functional and input operators
$\hat{Q}_{k}$ and $\hat{P}_{k}$ by a simple and physically transparent way.
This makes SRPA very convenient for the analysis and treatment of the
obtained results.
\\
$\bullet$
It is instructive to express the basic SRPA operators via
the separable residual interaction (\ref{eq:V_res_sep}):
\begin{equation} \label{eq:XY_V_res}
\hat{X}_{k} = [\hat{V}_{res}, \hat{P}_k]_{ph}, \qquad
\hat{Y}_{k} = [\hat{V}_{res}, \hat{Q}_k]_{ph}
\end{equation}
where the index $ph$ means the $1ph$ part of the operator. It is seen that
the T-odd operator $\hat{P}_k$ retains the T-even part of $V_{res}$
to build $\hat{X}_{k}$. Vice versa, the commutator with the T-even operator
$\hat{Q}_k$ keeps the T-odd part of $V_{res}$ to build $\hat{Y}_{k}$.
\\
$\bullet$
Some of the SRPA values read as averaged commutators between
T-odd and T-even operators. This allows to establish useful relations
with other models. For example, (\ref{eq:kappa}), (\ref{eq:eta}) and
(\ref{eq:XY_V_res}) give
\begin{eqnarray}
\label{eq:kappa_V}
\kappa_{k'k}^{-1 } = - i <0|[\hat{P}_{k'},{\hat X}_{k}]|0> =
- i <0|[\hat{P}_{k'},[\hat{V}_{res},{\hat P}_{k}]]|0> ,
\\
\label{eq:eta_V}
\eta_{k'k}^{-1 } = -i <0|[\hat{Q}_{k'},{\hat Y}_{k}]|0> =
-i <0|[\hat{Q}_{k'},[\hat{V}_{res},{\hat Q}_{k}]]|0> .
\end{eqnarray}
The similar double commutators but with the full Hamiltonian (instead
of the residual interaction) correspond to $m_3$ and $m_1$ sum rules,
respectively, and so represent the spring and inertia parameters
\cite{Re_PRA_90} in the basis of collective generators $\hat{Q}_k$
and $\hat{P}_k$. This allows to establish the connection of the SRPA with
the sum rule approach \cite{BLM_79,LS_89} and local RPA \cite{Re_PRA_90}.

Besides, the commutator form of  SRPA values can considerably simplify
their calculation (see discussion in the Appendix B).
\\
$\bullet$
 SRPA restores the conservation laws (e.g. translational invariance)
violated by the static mean field. Indeed, let's assume a symmetry mode
with the generator $\hat{P}_{\rm sym}$. Then, to keep the conservation law
$[\hat{H},\hat{P}_{\rm sym}]=0$,  we simply have to include
$\hat{P}_{\rm sym}$ into the set of the input generators $\hat{P}_k$
together with its complement $\hat{Q}_{\rm sym}=i[\hat{H},\hat{P}_{\rm sym}]$.
\\
$\bullet$ SRPA equations are very general and can be applied to diverse systems
(atomic nuclei, atomic clusters, etc.) described by density and
current-dependent functionals. Even Bose systems can be covered if to redefine
the many-body wave function (\ref{Psi_Thouless}) exhibiting the perturbation
through the elementary excitations. In this case, the Slater deterninant for
1ph excitations should be replaced by a perturbed many-body function in terms
of elementary bosonic excitations.
\\
$\bullet$ In fact, SRPA is the first TDLDA iteration with the initial wave
function (\ref{eq:scaling}). A single iteration is generally not enough to get the
complete convergence of TDLDA results. However, SRPA calculations demonstrate
that high accuracy can be achieved even in this case if to ensure the optimal
choice of the input operators  $\hat{Q}_{k}$ and
$\hat{P}_{k}$ and keep sufficient amount of the separable terms (see
discussion in Sec. 5). In this case, the first iteration already gives quite
accurate results.
\\
$\bullet$
There are some alternative RPA schemes also delivering self-consistent
factorization of the two-body residual interaction, see e.g.
\cite{LS_nuclei,SS,Kubo,Vor} for atomic nuclei and \cite{YB_92,babst} for
atomic clusters. However, these schemes are usually not sufficiently
general. Some of them are limited to analytic or simple numerical estimates
\cite{LS_nuclei,SS,YB_92}, next ones start from phenomenological
single-particle potentials
and thus are not fully self-consistent \cite{Kubo}, the others need a large number
of the separable terms to get an appropriate  numerical accuracy \cite{Vor,babst}.
SRPA has evident advantages as compared with these schemes.
\\
$\bullet$ After solution of the SRPA problem, the Hamiltonian
(\ref{eq:H_sep})-(\ref{eq:V_res_sep}) is reduced to a composition of
one-phonon RPA excitations
\begin{equation}
\hat{H}=\sum_{\nu} \omega_{\nu} \hat{Q}^+_{\nu}\hat{Q}_{\nu}
\end{equation}
where one-phonon operators are given by $(\ref{eq:Q_op})-(\ref{eq:Q_norm})$.
Then, it is easy to get expressions of the equation-of-motion (EOM) method:
\begin{equation}
 [\hat{H},\hat{Q}^+_{\nu}]=\omega_{\nu}\hat{Q}^+_{\nu} , \quad
[\hat{H},\hat{Q}_{\nu}]= - \omega_{\nu}\hat{Q}_{\nu} .
\end{equation}
So, SRPA and EOM with the Hamiltonian (\ref{eq:H_sep})-(\ref{eq:V_res_sep})
are equivalent. This allows to to establish
the connection between the SRPA and couled-cluster EOM method with the single reference
(see for reviews
\cite{Barlett_99,Paldus_99,Piecuch_02,Piecuch_03}). SRPA uses the excitation
operators involving only singles (1ph) and so generally carries less correlations
than EOM-CC. At the same time,  SRPA delivers very elegant and physically
transparent calculations scheme and, as is shown in our calculations, the
correlations included to the SRPA are often quite enough to describe linear dynamics.
It would be interesting to construct the approach combining
advantages of SRPA and EOM-CC.

\section{Strength function}

In study of response of a system to external fields, we are usually
interested in the average strength function instead of the responses
of particular RPA states. For example, giant resonances in heavy nuclei
are formed by thousands of RPA states whose contributions in any case cannot
be distinguished experimentally. In this case, it is reasonable
to consider the averaged response described by the strength function.
Besides, the calculation of the strength function is usually much easier.

For electric external fields of multipolarity $E\lambda\mu$, the strength
function can be defined as
\begin{equation}\label{eq:strength_function}
S_L(E\lambda\mu ; \omega)= \sum_{\nu}
\omega_{\nu}^{L} M_{\lambda\mu \nu}^2 \zeta(\omega - \omega_{\nu})
\end{equation}
where
\begin{equation}
\zeta(\omega - \omega_j) = \frac{1}{2\pi}
  \frac{\Delta}{(\omega - \omega_{\nu})^2 + (\Delta/2)^2}              
\label{eq:lorfold}
\end{equation}
is Lorentz weight with an averaging parameter $\Delta$
and
\begin{equation}\label{eq:tr_me}
M_{\lambda\mu \nu} = \frac{1}{2} \sum_{ph}
<ph|\hat{f}_{\lambda\mu}|0> ( c^{\nu -}_{ph} + c^{\nu +}_{ph}) \,
\end{equation}
is the transition matrix element for the external field
\begin{equation}
\hat{f}_{\lambda\mu} =
\frac{1}{1+\delta_{\mu ,0}}
r^{\lambda} (Y_{\lambda \mu} +  Y^{\dag}_{\lambda \mu}) .
\end{equation}
It is worth noting that, unlike the standard definition of the strength
function with using $\delta (\omega - \omega_{\nu})$, we exploit here
the Lorentz weight. It is very convenient to simulate various smoothing effects.

The explicite expression for (\ref{eq:strength_function}) can be obtained by using
the Cauchy residue theorem. For this aim, the strength function is recasted as a sum
of $\nu$ residues for the poles $z=\pm \omega_{\nu}$. Since the sum of all the
residues (covering all the poles) is zero, the residues with $z=\pm
\omega_{\nu}$ (whose calculation is time consuming) can be replaced by the sum
of residies with  $z=\omega \pm i(\Delta /2)$ and $z=\pm \varepsilon_{ph}$ whose
calculation is much less expensive (see details of the derivation in
\cite{Ne_AP_02}).

Finally, the strength function for L=0 and 1 reads as
\begin{eqnarray}
\label{eq:sf}
S_L (E\lambda\mu ,\omega )
&=&\frac{1}{\pi} \Im
\left[
\frac{z^{L}\det|B(z)|}{\det|F(z)|}
\right]_{z=\omega +i(\Delta /2)}
\\
&+& 2\sqrt{2}
\sum_{ph}^{K_p,K_h>0}
\varepsilon_{ph}^{L} <ph|\hat{f}_{\lambda\mu}|0>^2
\zeta(\omega -\varepsilon_{ph}) .
\nonumber
\end{eqnarray}
 The first term in (\ref{eq:sf})
is contribution of the residual two-body interaction while the second term is the unperturbed
(purely $1ph$) strength function. Further, $F(z)$ is determinant of the RPA symmetric matrix
(\ref{eq:RPA_eq}) of the rank $2K$, where $K$ is the number of the initial operators $\hat{Q_k}$.
 The symmetric matrix $B$ of the rank $(2K+1)$ is defined as
\begin{eqnarray}\label{eq:B_matrix}
B_{nn'}(z)&=&F_{nn'}(z),
\\
B_{2K+1,2K+1}(z)&=&0, \quad B_{2K+1,n}(z)=B_{n,2K+1}(z)=A_{n}(z)
\nonumber
\end{eqnarray}
where $n,n'=1, ... , 2K$ and left (right) indexes define lines (columns).

The values $A_n(z)$ read
\begin{eqnarray}\label{eq:A_X}
A_{n=2k-1}^{(X)}(E\lambda\mu, z) &=&
4\sum_{ph}^{K_p,K_h>0}
\frac{\varepsilon_{ph}<ph|\hat{X}_{k}|0>
<ph|\hat{f}_{\lambda\mu}|0>}
      {\varepsilon_{ph}^2-z^2}
\\
A_{n=2k}^{(Y)}(E\lambda\mu, z) &=&
4\sum_{ph}^{K_p,K_h>0}
\frac{z{<ph|\hat{Y}_{k}|0><ph|\hat{f}_{\lambda\mu}|0>}}
      {\varepsilon_{ph}^2-z^2}  .
\nonumber
\end{eqnarray}
They form the right and low borders of the determinant $B$ thus fringering
the RPA determinant $F$. The values $-A_n(z)$ in the most right column
have the same indices as the corresponding strings of the RPA determinant.
The values $A_n(z)$ in the lowest line
have the same indices as the corresponding columns of the RPA determinant.

Derivation of the strentth function, given above, deviates from the standard one in
the lineary response theory. Besides, the SRPA deals with the Lorentz weight instead
of $\delta (\omega - \omega_{\nu})$ used in the linear response theory. At the same
time, SRPA strength function and lineary response theory are conceptually
the same approaches. Since the linear response theory is widely used
in the coupled-cluster (CC) method, it would be interesting to consider
the implementation of SRPA stength function method to CC.
The linear response theory is widely used
in the coupled-cluster (CC) method. In this connection, it would be interesting to
enlarge the SRPA stength function method to CC.

\section{Particular cases for clusters and nuclei}
\label{sec:KS_Skyrme}
\subsection{Kohn-Sham functional for atomic clusters}

Kohn-Sham functional for atomic clusters reads
\begin{equation} \label{eq:func}
E_{\rm tot}(t)= E_{\rm kin}(t) + E_{xc}(t) + E_C(t)=\int d{\vec r} {\cal H} (\rho({\vec r},t))
\end{equation}
where
\begin{eqnarray}
  E_{\rm kin}(t)
  &=&
  \frac{\hbar^2}{2m_e} \int d{\vec r} \, \tau ({\vec r},t)
  \;,
\label{eq: E_T}\\
  E_{xc}(t)
  &=&
  \int d{\vec r}\,\rho({\vec r},t)\,
   \epsilon_{xc}(\rho\left({\vec r},t)\right)
  \;,
\label{eq:exc} \\
  E_C(t)
  &=&
  \frac{e^2}{2}\int \int  d{\vec r}d{\vec r}_1
  \frac{(\rho({\vec r},t) - \rho_i({\vec r}))
        (\rho({\vec r}_1,t) - \rho_i({\vec r}_1))}
       {\vert {\vec r} - {\vec r}_1\vert}
\label{eq:ec}
\end{eqnarray}
\noindent
are kinetic, exchange-correlation (in the local density approximation),
and Coulomb terms, respectively. Further, $\rho_i({\vec r})$ is the ionic
density and $\rho({\vec r},t)$ and $\tau ({\vec r},t)$
are density and kinetic energy density of valence
electrons.

In atomic clusters, oscillations of valence electrons are generated by time-dependent
variations of the electronic T-even density $\rho({\vec r},t)$ only. So, one may
neglect in the SRPA formalism all T-odd densities and their variations $p_{k}(t)$.
This makes SRPA equations especially simple. In particular,
the density variation (\ref{eq:trans_dens})  is reduced to
\begin{eqnarray}\label{eq:KS_trans_dens}
\delta \rho ({\vec r},t)&&  =
i \sum_k  q_{k}(t)<0|[\hat{P}_{k},\hat \rho (\vec r)]|0>
\\
&& =
-4i \sum_k  q_{k}(t) \sum_{ph}^{K_p,K_h>0}
<ph|\hat{P_k}|0> \Re{<ph|\hat{\rho}|0>}
\nonumber\\
&&  = i \sum_k q_{k}(t) \delta \rho ({\vec r})
 \nonumber
\end{eqnarray}
where
\begin{equation}\label{eq:<ph|P0>}
<ph|\hat{P}_k|0> = 2i \varepsilon_{ph} <ph|\hat{Q}_k|> ,
\end{equation}
\begin{equation}\label{rho_r}
\delta \rho ({\vec r})= -\frac{\hbar^2}{2m_e}
({\vec \bigtriangledown} \bar{\rho}({\vec r}) \cdot
 {\vec \bigtriangledown} Q_{k}({\vec r}) + 2 \bar{\rho} ({\vec r})
 \triangle Q_{k}({\vec r})) .
\end{equation}
Here, $\Re{<ph|\hat{\rho}|0>}$ is the transition density  and
$\bar{\rho}({\vec r})$ is the static ground state density of valence electrons.

It is seen from (\ref{eq:KS_trans_dens})-(\ref{eq:<ph|P0>}) that there are two
alternative ways to calculate the  density variation: i) through the transition
density and matrix elements of $\hat{Q}_k$-operator and ii) through the ground
state density. The second way is the most simple. It becomes possible because,
in atomic clusters, $V_{res}$ has no T-odd $\hat{Y}_k$-operators and thus the
commutator of $\hat{Q}_k$ with the full Hamiltonian is reduced to the
commutator with the kinetic energy term only:
\begin{equation}\label{eq:P}
\hat{P}_{k} = i[\hat{H},\hat{Q}_{k}]_{ph} = i[\hat{h}_0,\hat{Q}_{k}]_{ph}
= -i\frac{\hbar^2}{2m_e}[\vec{\bigtriangledown}^2,\hat{Q}_{k}]_{ph} .
\end{equation}
This drastically simplifies SRPA expressions and allows to present them in terms
of the static ground state density. The scaling tranformation (\ref{eq:scaling})
loses exponents with  $p_{k}(t)$ and reads
\begin{equation}\label{eq:KS_scaling}
|\Psi(t)\!>=\prod_{k=1}^K
exp\{ -i q_{k}(t)\hat{P}_{k}\} |0> .
\end{equation}
Other SRPA equations are reduced to
\begin{eqnarray}\label{eq:X_KS}
\hat{X}_{k}({\vec r})   &=& i
[\frac{\delta^2 {\cal H}} {\delta \rho
\delta \rho}]_{\rho=\bar{\rho}}
<0|[\hat{P}_{k} ,{\hat \rho}]|0>
{\hat \rho}(\vec r)
\nonumber\\
&=& (\frac{\partial^2 {\cal H}_{xc}}{\partial\rho \partial\rho})_{\rho ={\bar \rho}}
\delta \rho ({\vec r})
+ e^2 \int d{\vec r}_1
\frac{\delta \rho ({\vec r}_1)}{\vert {\vec r} - {\vec r_1}\vert } ,
\end{eqnarray}
\begin{eqnarray}
\label{eq:KS_kappa}
\kappa_{k'k}^{-1 }&&=
\kappa_{kk'}^{-1} =
- i <0|[\hat{P}_{k'},{\hat X}_{k}]|0>
\nonumber\\
&& = \int d{\vec r}
\{ \frac{\delta^2 {\cal H}}{\delta \rho \delta \rho}\}_{\rho ={\bar \rho}}
<0|[\hat{P}_{k},{\hat \rho}]|0>
<0|[\hat{P}_{k'},{\hat \rho}]|0>
\nonumber \\
&& =
 - \int d{\vec r} X_{k}({\vec r})\delta \rho ({\vec r}) ,
\end{eqnarray}
 \begin{equation}
\label{eq:KS_RPA_eq}
\sum_{k} \bar{q}_{k}^{\nu}
\{ F_{k'k}^{(XX)}- \kappa_{kk'}^{-1}] \}=0 ,
\end{equation}
\begin{equation}
c^{\nu\pm}_{ph}=-\frac{1}{2}
\frac{\sum_{k} \bar{q}_{k}^{\nu}<ph|\hat{X}_{k}|0>}
{\varepsilon_{ph}\pm\omega_{\nu}},
\label{eq:KS_c_pm_qp}
\end{equation}
\begin{equation}\label{eq:KS_c_norm_F}
  \sum_{ph} \{ (c^{\nu -}_{ph})^2 - (c^{\nu +}_{ph})^2 \}
=  \sum_{kk'}\frac{1}{4}
\bar{q}^{\nu}_{k} \bar{q}_{k'}^{\nu}
\frac{\partial F_{kk'}^{(XX)}(\omega_{\nu})}{\partial\omega_{\nu}}
= 2 N_{\nu} .
\end{equation}
It is seen that the basic operator (\ref{eq:X_KS}) and strength matrix
(\ref{eq:KS_kappa}) have now simple expressions via
$\delta \rho ({\vec r})$ from (\ref{rho_r}). The operator (\ref{eq:X_KS})
has exchange-correlation and Coulomb terms. For electric multipole oscillations
(dipole plasmon, ...), the Coulomb term dominates.

In recent years, there appear some new functionals where the current of electrons
instead of their density is used as a basic variable \cite{KS_current}. SRPA
equations for this case can be straightforwardly obtained from the general formalism
given in Sec. 2.

\subsection{Skyrme functional for atomic nuclei}

Nuclear interaction is very complicated and its explicit form is still unknown.
So, in practice different approximations to nuclear interaction are used. Skyrme
forces \cite{Skyrme,Engel_75} represent one of the  most successful approximations
where the interaction is maximally simplified and, at the same time, allows to get
accurate and universal description of both ground state properties and dynamics of
atomic nuclei (see \cite{Ben_RMP_03} a for recent review). Skyrme forces are
contact, i.e.  $\sim \delta ({\vec r}_1-{\vec r}_2)$, which minimizes the computational
effort. In spite of this dramatic simplification, Skyrme forcese well
reproduce properties of most spherical and deformed nuclei  as well as characteristics
of nuclear matter and neutron stars. Additional advantage of the Skyrme
 interaction is that its parameters are directly related to the basic nuclear properties:
 incompressibility, nuclear radii, masses and binding energies, etc. SRPA for Skyrme forces
 was derived in \cite{Miori_01,Ne_PRC_1d,Prague_02,LS}.

Although Skyrme forces are relatively simple, they are still
much more demanding than the Coulomb interaction. In particular, they deal with a variety
of diverse densities and currents.
The Skyrme functional reads \cite{Engel_75,Re_nucl_AP_92,Dob_PRC_95}
\begin{equation}
   {E}  =  \int d{\vec r}\left({\cal H}_{\rm kin}
 +{\cal H}_{\rm Sk}(\rho_s,\tau_s,
   \vec{\sigma}_s,\vec{j}_s,\vec{J}_s)
                           + {\cal H}_{\rm C}(\rho_p) \right) ,
\end{equation}
where
\begin{eqnarray}
   {\cal H}_{\rm kin} &= &  \frac{\hbar^2}{2m} \tau ,
 \label{Ekin}
 \\
 {\cal H}_{\rm C}
 & = & \frac{e^2}{2}
 \int  d{\vec r}' \rho_p(\vec{r})
              \frac{1}{|\vec{r}-\vec{r}'|} \rho_p(\vec{r}')
         -\frac{3}{4} e^2(\frac{3}{\pi})^\frac{1}{3}
                          [ \rho_p(\vec{r})]^\frac{4}{3} ,
\label{Ecoul}
\\
   {\cal H}_{\rm Sk} &= &
                 \frac{b_0}{2}  \rho^2
                  -\frac{b'_0}{2} \sum_s \rho_s^2
    -\frac{b_2}{2} \rho \Delta \rho
     +\frac{b'_2}{2} \sum_s \rho_s \Delta \rho_s
    \\
        & &
          + \frac{b_3}{3}  \rho^{\alpha +2}
          - \frac{b'_3}{3} \rho^\alpha   \sum_s\rho_s^2
            \nonumber \\
     & &
     +  b_1 (\rho \tau - \vec{j}^2)
     - b'_1 \sum_s (\rho_s \tau_s - \vec{j}_s^2)
\nonumber\\
     & &
     - b_4 \left( \rho\vec{\nabla}\vec{{\Im}}
      + \vec{\sigma} \cdot (\vec{\nabla} \times \vec{j})\right)
     -b'_4 \sum_s \left( \rho_s(\vec{\nabla} \vec{\Im}_s)
              + \vec{\sigma}_s \cdot (\vec{\nabla} \times \vec{j}_s) \right)
\nonumber\\
     & &
      + \tilde{b}_4\left( \vec{\sigma}\vec{T}-\vec{\Im}^2 \right)
      + \tilde{b}'_4\sum_s \left( \vec{\sigma}_s\vec{T}_s-\vec{\Im}_s^2 \right)
 \nonumber \label{Esky}
\end{eqnarray}
are kinetic, Coulomb and Skyrme terms respectively. The isospin index  $s=n,p$
covers neutrons (n) and protons (p).
Densities without this index involve both neutrons and protons,
e.g. $\rho=\rho_p+\rho_n$. Parameters $b$ and $\alpha$ are fitted
to describe ground state properties of atomic nuclei.

The functional  includes diverse densities and currents, both neutrons
and protons. They are naturally separated into two groups: 1) T-even density
$\rho_s(\vec{r})$, kinetic energy density
$\tau_s(\vec{r})$ and spin orbital density $\vec{\Im}_s(\vec{r})$ and 2) T-odd
spin density $\sigma_s(\vec{r})$, current $\vec{j}_s(\vec{r})$ and vector kinetic
energy density $\vec{T}_s(\vec{r})$. Explicit
expressions for these densities and currents, as well as for their operators,
are given in the appendix A. Only T-even densities contribute to the ground
state properties of nuclei with even numbers of protons and neutrons (and thus
with T-even wave function of the ground state). Instead, both
T-even and T-odd densities participate in generation of nuclear oscillations.
Sec. 2 delivers the SRPA formalism for this general case.

\subsection{T-odd densities and currents}

 Comparison of the Kohn-Sham and Skyrme functionals leads to  a natural question
why these two functionals exploit, for the time-dependent problem, so different
sets of basic densities and currents? If the Kohn-Sham functional is content with one
density, the Skyrme forces operate with a diverse set of densities and currents,
both T-even and T-odd. Then, should we consider T-odd densities as
genuine for the description of dynamics of finite many-body systems or they are
a pequliarity of nuclear forces? This question is very nontrivial and still poorly
studied. We present below some comments which, at least partly, clarify this point.

 Actual nuclear forces are of a finite range. These  are, for example, Gogny
forces \cite{Gogny_75} representing more realistic approximation of actual
nuclear forces than Skyrme approximation. Gogny interaction has no any velocity
dependence and fulfills the Galilean invariance. Instead, two-body Skyrme
interaction depends on relative velocities $\vec{k}=1/2i \cdot (\vec{\nabla}_1
- \vec{\nabla}_2)$, which just simulates the finite range effects \cite{Ben_RMP_03}.

 The static Hartree-Fock problem assumes T-reversal invariance and T-even
single-particle density matrix. In this case, Skyrme forces can be limited by
only T-even densities: $\rho_s(\vec{r})$, $\tau_s(\vec{r})$ and  $\vec{\Im}_s(\vec{r})$.
In the case of dynamics, the density matrix is not already T-even and aquires
T-odd components \cite{Engel_75}.  This fact,  together with velocity dependence
of the Skyrme interaction, results in appearance in the Skyrme functional
of T-odd densities and currents: $ \vec{s}_s(\vec{r})$, $\vec{j}_s(\vec{r})$ and
$\vec{T}_s(\vec{r})$ \cite{Engel_75,Re_nucl_AP_92}. Hence the origin of T-odd
densities in the Skyrme functional. However, this is not the general case for nuclear
forces.

 As compared with the Kohn-Sham functional for electronic systems, the nuclear
Skyrme functional is less genuine. The main (Coulomb) interaction in the
Kohn-Sham problem is well known and only exchange and corellations should be modeled.
Instead, in the nuclear case, even the basic interaction is unknown and should
be approximated, e.g. by the simple contact interaction in Skyrme forces.

The crudeness of Skyrme forces has certain consequences. For example,
the Skyrme functional has no any exchange-correlation term  since the
relevant effects are supposed to be already included into numerous Skyrme fitting
parameters. Besides, the Skyrme functional may accept a diverse set of T-even and T-dd
densities and currents. One may say that T-odd densities appear in the Skyrme
functional partly because of its specific construction. Indeed, other effective
nuclear forces (Gogny \cite{Gogny_75} , Landau-Migdal \cite{LM}) do not exploit
T-odd densities and currents for description of nuclear dynamics.

Implementation of a variety of densities and currents in the Skyrme fuctional has,
however,
some advantages. It is known that different projectiles and external fields
used to generate collective modes in nuclear reactions are often selective
to particular densities and currents. For example,
some elastic magnetic collective modes (scissors, twist) are associated with variations
in the momentum space while keeping the common density $\rho_s(\vec{r})$  about constant.
T-odd densities and currents can play here a significant role while
Skyrme forces obtain the advantage to describe the collective motion by a natural
and physically transparent way.

Relative contributions of T-odd densities to a given mode should obviously
depend on the character of this mode. Electric multipole excitations (plasmons
in atomic clusters, $E\lambda$ giant resonances in atomic nuclei) are mainly provided
 by T-even densities (see e.g. \cite{Prague_02}). Instead, T-odd
densities and currents might be important for magnetic modes and
maybe some exotic (toroidal, ...) electric modes.

It worth noting that T-odd denstities appear in the Skyrme functional in the
specific combinations $\rho_s \tau_s - \vec{j}_s^2$,
$\rho_s(\vec{\nabla} \vec{\Im}_s)
              + \vec{\sigma}_s \cdot (\vec{\nabla} \times \vec{j}_s)$,
and $\vec{\sigma}_s\vec{T}_s-\vec{\Im}_s^2$.
 Following \cite{Dob_PRC_95}, this ensures Skyrme forces
to fulfill the local gauge invariance (and Galilean invariance as the particular
case). The velocity-independent finite-range Gogny forces also keep this invariance.
Being combined into specific combinations, T-odd densities do not require any new Skyrme
parameters \cite{Dob_PRC_95}. So, the parameters fitted to the static nuclear properties
with T-even densities only, are enough for description of the dynamics as well.

The time-dependent density functional theory \cite{RG_84} for electronic systems
is usually implemented at adiabatic local density approximation (ALDA)
when density and single-particle potential are supposed to vary slowly both
in time and space. Last years, the current-dependent Kohn-Sham functionals with a
current density as a basic variable were introduced to treat the collective motion beyond
ALDA (see e.g. \cite{KS_current}). These functionals are robust
for a time-dependent linear response problem where the ordinary density
functionals become strongly nonlocal. The theory is reformulated
in terms of a vector potential for exchange and correlations,
depending on the induced {\it current} density. So,  T-odd variables
appear in electronic functionals as well.

In general, the role of T-odd variables in dynamics of finite many-body systems
is still rather vague. This fundamental problem devotes deep and comprehensive
study.

\section{Choice of initial operators}
\label{sec:choice}

It is easy to see that, after choosing  the initial operators
$\hat{Q}_{k}({\vec r})$, all other SRPA values can be straightforwardly
determined following the steps
$$
\hat{Q}_k \; \Rightarrow \;
<|[\hat{Q}_k,{\hat J}_{\alpha}]|> \; \Rightarrow \;
\hat{Y}_k, \; \eta_{kk'}^{-1} \; \Rightarrow \;
\hat{P}_k \; \Rightarrow \;
<|[\hat{P}_k,{\hat J}_{\alpha}]|> \; \Rightarrow \;
\hat{X}_k, \; \kappa_{kk'}^{-1} .
$$
As was mentioned above, the proper choice of initial operators
$\hat{Q}_{k}({\vec r})$ is crucial to achieve good convergence of the
separable expansion (\ref{two_body}) with a minimal number of separable
operators.

SRPA itself does not give a recipe to determine $\hat{Q}_{k}({\vec r})$.
But choice of these operators can be  inspired by physical and computational
arguments. The operators  should be simple and universal in the sense that they can
be applied equally well to all modes and excitation channels. The main
idea is that the initial  operators should result in exploration of different spatial
regions of the system, the surface and interior.  This
suggests that the leading scaling operator should have the form of the
applied external field in the long-wave approximation, for example,
\begin{equation} \label{Q_lm}
\hat{Q}^{\lambda\mu}_{k=1}({\vec r})=r^{\lambda}
(Y_{\lambda\mu}(\Omega)+\mbox{h.c.)} .
\end{equation}
Such a choice results in the
separable operators (\ref{eq:X}), (\ref{eq:Y}) and (\ref{eq:X_KS})
most sensitive to the surface of the system. This is evident
in (\ref{eq:X_KS}) where $\delta \rho ({\vec r}) \propto
{\vec \bigtriangledown} \bar{\rho}({\vec r})$ is peaked at the surface.
Many collective oscillations manifest themselves as predominantly surface modes.
As a result, already one separable term generating by (\ref{Q_lm}) usually delivers
a quite good description of collective excitations like plasmons in atomic
clusters and giant resonances in atomic nuclei. The detailed distributions
depends on a subtle interplay of surface and volume vibrations.  This
can be resolved by taking into account the nuclear interior. For
this aim, the radial parts with larger powers
$r^{\lambda+n_k}Y_{\lambda\mu}$ and spherical Bessel functions can be
used, much similar as in the local RPA
\cite{Re_PRA_90}. This results in the shift of the maxima
of the operators (\ref{eq:X}), (\ref{eq:Y}) and (\ref{eq:X_KS}) to the interior.
Exploring different conceivable combinations,
one may found a most efficient set of the initial operators.

\begin{figure}[h]
\includegraphics [height=10cm,width=8cm]{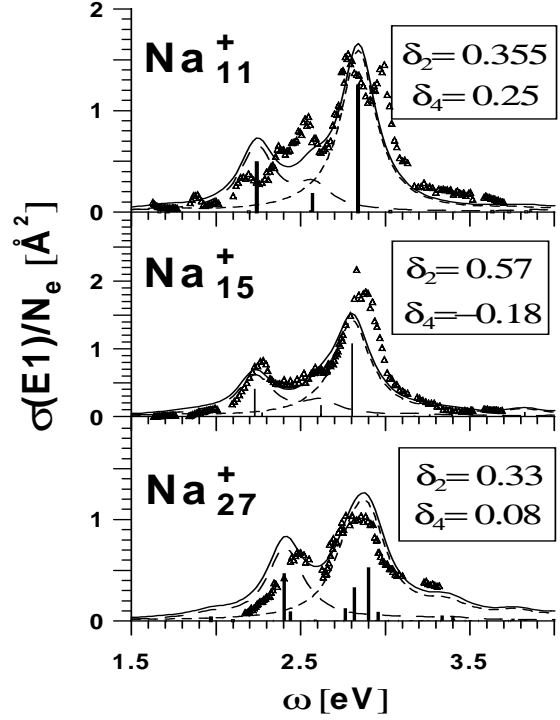}
\caption{\label{fig:e1_na}
Photoabsorption cross section for the dipole plasmon in axially deformed
sodium clusters, normalized to the number of valence electrons $N_e$. The
parameters of quadrupole and hexadecapole deformations are given in boxes.
The experimental data \protect\cite{SH} (triangles) are compared with SRPA
results given as bars for RPA states  and as the strength function
(\protect\ref{eq:strength_function}) smoothed by the Lorentz weight
with $\Delta = 0.25$ eV. Contributions to the strength function from
$\mu =$0 and 1 dipole modes (the latter has twice larger strength) are exhibited
by dashed curves. The bars are given in $eV \AA^2$.
}
\end{figure}

For description of the dipole plasmon in atomic clusters, the set of
hermitian operators
\begin{equation} \label{eq:lop}
Q^{\lambda_k\mu}_k ({\vec r})=r^{\lambda_k + n_k}
(Y_{\lambda_k\mu}(\Omega)+                                  
Y_{\lambda_k\mu}^{\dag}(\Omega))
\end{equation}
with $\lambda_k n_k$ = 10, 12, 14  and $\mu =0,1$ is usually enough
\cite{Ne_AP_02,Ne_EPJD_02}.
As is seen from Fig. 1, we successfully reproduce gross structure of the dipole
plasmon in light axially-deformed sodium clusters (some discrepancies for
the lightest cluster  Na$_{11}^+$ arise because of the roughness of the ionic jellium
approximation for smallest samples). Already one initial operator is usually enough
to reproduce the energy of the dipole plasmon and its branches but in this case
the plasmon acquires some artificial strength in its right flank and thus the
overestimated width \cite{Ne_PRA_98}. This problem can be solved by adding two more
initial operators.
The calculations for a variety of spherical alkali metal clusters \cite{Ne_EPJD_98}
as well as for deformed clusters of a medium size \cite{Ne_EPJD_02} show
that SRPA correcly describes not only gross structure of the dipole plasmon
but also its Landau damping and width.

For the description of giant resonances in atomic nuclei,
we used the set of initial operators \cite{Ne_PRC_1d}
\begin{equation}
  \hat{Q}_{k}({\vec r})
  =
  R_k(r)(Y_{\lambda\mu} (\Omega )+\mbox{h.c.})
\label{eq:scale_op}
\end{equation}
with
\begin{equation}
  R_k(r)
  =
  \left\{
  \begin{array}{ll}
   r^{\lambda }, & k\!=\!1
  \\
  j_{\lambda}(q^k_{\lambda}r), & k\!=\!2,3,4
\\
\end{array}
\right.
\label{eq:actualset}
\end{equation}
$$
  q^{k}_{\lambda} = a_k\frac{z_{\lambda}}{R_{\rm diff}} ,\quad
  a_2\!=\!0.6\;,
  a_3\!=\!0.9\;,
  a_4\!=\!1.2
$$
where $R_{\rm diff}$ is the diffraction radius of the actual nucleus
and $z_{\lambda}$ is the first root in of the spherical Bessel function
$j_{\lambda}(z_{\lambda})=0$. The separable term with $k=1$ is
mainly localized at the nuclear surface while the next terms are
localized more and more in the interior.  This simple set seems to be
a best compromise for the description of nuclear giant resonances in
light and heavy nuclei.
\begin{figure}[t]
\begin{center}
\includegraphics[height=5cm]{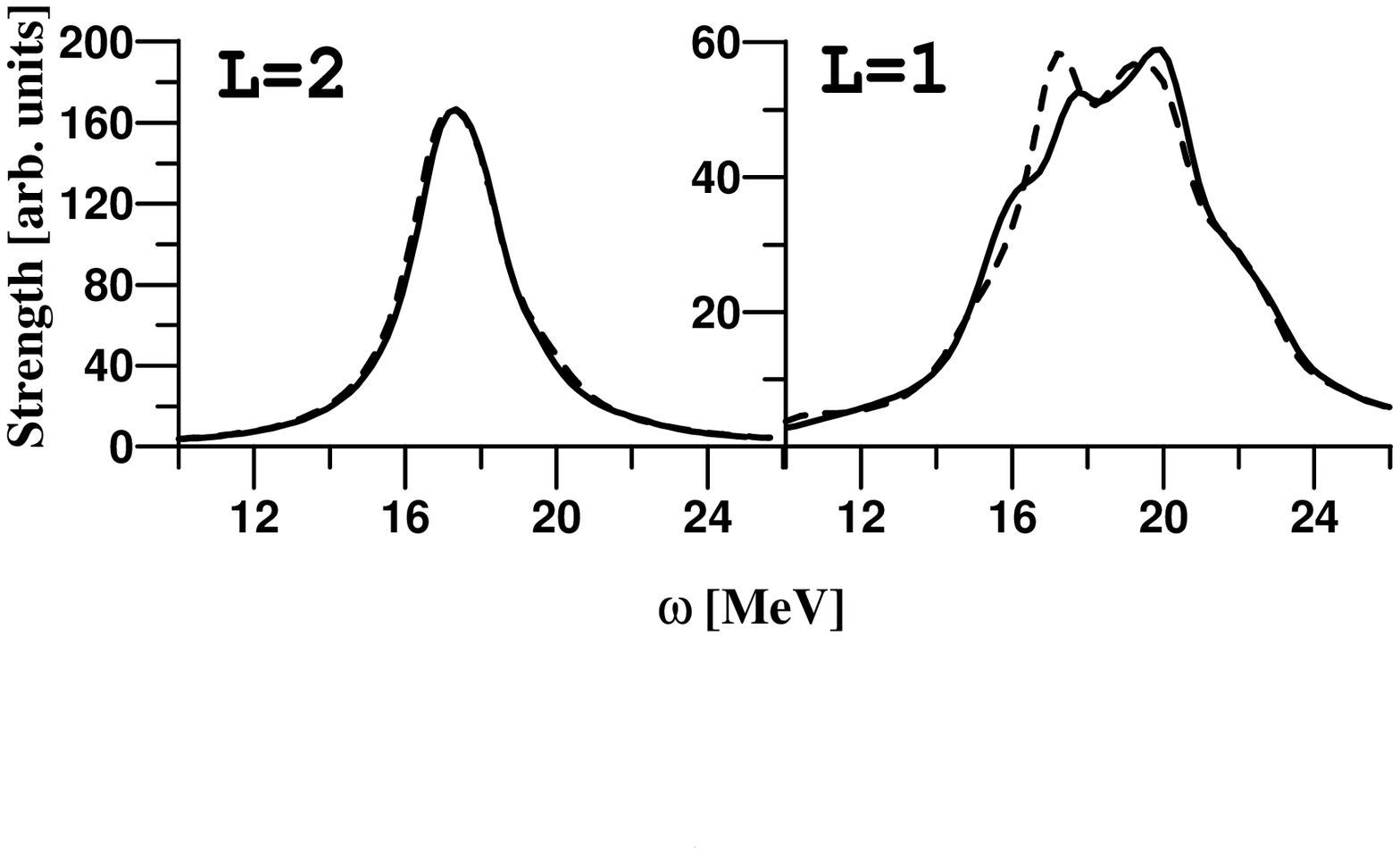}
\end{center}
\caption
{Isoscalar E2 and isovector E1 giant resonances in $^{40}Ca$ calculated with
SkM* forces. The results are exhibited for full (exact) RPA (solid curve) and SRPA
with $k=1$ (dotted curve).
}
\label{fig:comp_srpa}
\end{figure}
Fig. 2 demonstrates that already one separable term (k=1) can be enough to get
a reasonable agreement with the exact results. For k=1, the calculations are
especially simple and results are easily analyzed.

The sets (\ref{eq:lop})-(\ref{eq:actualset}) are optimal for description of
electric collective modes ($E\lambda$ plasmons in clusters and giant
resonances in nuclei). For description of magnetic modes, the initial
operator should resemble the T-odd magnetic external field. So, in this case
we should start from the initial operators $\hat{P}_k$ in the form of the
magnetic multipole transition operator in the long-wave approximation. The T-even
operators $\hat{Q}_k$ are then obtained from the connection
$\hat{Q}_k=i[\hat{H},\hat{P}_k]$.

\section{Summary}
\label{sec:summary}

We presented fully self-consistent separable random-phase-approximation
(SRPA) method for description of linear dynamics of different finite
Fermi-systems. The method is very general, physically transparent,
convenient for the analysis and treatment of the results. SRPA
drastically simplifies the calculations. It allows to get a high
numerical accuracy with a minimal computational effort. The method is
especially effective for systems with a number of particles $10-10^3$,
where quantum-shell  effects in the spectra and responses
are significant. In such systems, the familiar macroscopic methods are
too rough while the full-scale microscopic methods are too
expensive. SRPA seems to be here the best compromise between quality of the
results and the computational effort. As the most involved methods,  SRPA describes
the Landau damping, one of the most important characteristics of the
collective motion. SRPA results can be obtained in terms of both separate
RPA states and the strength function (linear response to external fields).

The particular SRPA versions for electronic Kohn-Sham and nuclear Skyrme functional
were considered and examples of the calculations for the dipole plasmon in atomic
clusters and giant resonances in atomic nuclei were presented. SRPA was compared with
alternative methods, in particular with EOM-CC. It would be interesting to combine
advantages of SRPA and couled-cluster approach in one powerful method.

\vspace{0.2cm} {\bf Acknowledgments.}
The work was partly supported  by the  DFG grant (project GZ:436 RUS 17/104/05)
and the grants
Heisenberg-Landau  (Germany - BLTP JINR) and Votruba-Blokhintcev
(Czech Republic - BLTP JINR).

\section*{Appendix A: Densities and currents for Skyrme functional}
\label{sec:dens_curr}
In Skyrme forces, the complete set of the densities involves the ordinary density,
kinetic-energy density, spin-orbital density, current density, spin density and vector
kinetic-energy density:
\begin{eqnarray*}
  \rho_s({\vec r},t)
  &=&
  \sum_{h \epsilon s}^{occ}
  \varphi^*_h({\vec r},t)\varphi_h^{\mbox{}}({\vec r},t) ,
  \qquad\qquad\qquad\qquad\qquad\qquad\quad
  \hat{T} \rho \hat{T}^{-1}=\rho
\\
  \tau_s({\vec r},t)
  &=&
  \sum_{h \epsilon s}^{occ}
  \vec{\nabla}\varphi^*_h({\vec r},t)\!\cdot\!
  \vec{\nabla}\varphi_h^{\mbox{}}({\vec r},t) ,
 \qquad\qquad\qquad\qquad\qquad\quad
  \hat{T} \tau \hat{T}^{-1}=\tau
\\
  \vec{\Im}_s({\vec r},t)
  &=&
  -i\sum_{h \epsilon s}^{occ}
  \varphi^*_h({\vec r},t)(\vec{\nabla}\times
  \hat{\vec{\sigma}})\varphi_h^{\mbox{}}({\vec r},t) ,
  \qquad\quad\qquad\qquad\quad
  \hat{T} {\vec \Im} \hat{T}^{-1}={\vec \Im}
\\
  \vec{j}_s({\vec r},t)
  &=&
  -\frac{i}{2}\sum_{h \epsilon s}^{occ}
  \left[
  \varphi^*_h({\vec r},t)\vec{\nabla}\varphi_h^{\mbox{}}({\vec r},t)
  -
  \vec{\nabla}\varphi^*_h({\vec r},t)\varphi_h^{\mbox{}}({\vec r},t)
  \right] ,
\quad
  \hat{T} {\vec j} \hat{T}^{-1}=-{\vec j}
\\
  \vec{\sigma}_s({\vec r})
  &=&
  \sum_{h \epsilon s}^{occ}
  \varphi^*_h({\vec r},t)\hat{\vec{\sigma}}
  \varphi_h^{\mbox{}}({\vec r},t) ,
\qquad\qquad\qquad\qquad\qquad\qquad \;\;
  \hat{T} {\vec \sigma} \hat{T}^{-1}=-{\vec \sigma}
\\
  \vec{T}_s({\vec r})
  &=&
  \sum_{h \epsilon s}^{occ}
  \vec{\nabla}\varphi^*_h({\vec r},t)\hat{\vec{\sigma}}
  \cdot \vec{\nabla}\varphi_h^{\mbox{}}({\vec r},t) ,
\qquad\qquad\qquad\qquad\qquad \;
  \hat{T} {\vec \sigma} \hat{T}^{-1}=-{\vec \sigma}
\end{eqnarray*}
where the sum runs over the occupied (hole) single-particle states $h$.
The  associated operators are
\begin{eqnarray*}
  \hat{\rho}_s(\vec{r})
  &=&
  \sum_{i=1}^{N_s}\delta(\vec{r}_i-\vec{r}) ,
\\
  \hat{\tau}_s(\vec{r})
  &=&
  \sum_{i=1}^{N_s}
  \overleftarrow{\nabla}\delta(\vec{r}_i - \vec{r})\vec{\nabla} ,
\\
  \hat{\vec{\Im}}_s(\vec{r})
  &=&
  \sum_{i=1}^{N_s}
  \delta(\vec{r_i} - \vec{r})\vec{\nabla}\!\times\!\hat{\vec{\sigma}} ,
\\
  \hat{\vec{j}}_s(\vec{r})
  &=&
  \frac{1}{2}\sum_{i=1}^{N_s}
  \left\{ \vec{\nabla},
  \delta(\vec{r}_i-\vec{r})
  \right\} ,
\\
  \hat{\vec{\sigma}}_s(\vec{r})
  &=&
  \sum_{i=1}^{N_s} \delta(\vec{r}_i-\vec{r})\hat{\vec{\sigma}} ,
\\
 \hat{\vec{T}}_s(\vec{r})
  &=&
  \sum_{i=1}^{N_s}
  \overleftarrow{\nabla}\delta(\vec{r}_i - \vec{r})\vec{\nabla} \hat{\vec{\sigma}} ,
\end{eqnarray*}
where $\hat{\vec{\sigma}}$ is the Pauli matrix, $N_s$ is number of protons or neutrons
in the nucleus.

\section*{Appendix B: Presentation of responses and strength matrices
through the matrix elements}
\label{sec:resp_strength_matr}

Responses $<0|[\hat{P}_{k} ,{\hat J}_{\alpha'}]|0>$ and
$<0|[\hat{Q}_{k} ,{\hat J}_{\alpha'}]|0>$ in (\ref{eq:X})-(\ref{eq:Y})
and inverse strength matrices in (\ref{eq:kappa})-(\ref{eq:eta})
read as the averaged commutators
\begin{equation}\label{eq:<[A,B]>}
  <0|[\hat{A},\hat{B}]|0> \quad \mbox{with} \quad
T\hat{A}T^{-1} = -T\hat{B}T^{-1} .
\end{equation}
Calculation  of these values can be greatly simplified if to express them
through the $1ph$ matrix elements of the operators $\hat{A}$ and $\hat{B}$.

In the case of the strength matrices, the matrix elements are
real for T-even operators and image for T-odd operators and
thus we easily get
\begin{eqnarray}
\label{eq:kappa_me}
\kappa_{k'k}^{-1 }=
- i <0|[\hat{P}_{k'},{\hat X}_{k}]|0>=
4i \sum_{ph}^{K_p , K_h>0}
<ph|\hat{P}_{k'}|0><ph|\hat{X}_{k}]|0> ,
\\
\label{eq:eta_me}
\eta_{k'k}^{-1 }= -i
<0|[\hat{Q}_{k'},{\hat Y}_{k}]|0>=
-4i \sum_{ph}^{K_p , K_h>0}
<ph|\hat{Q}_{k'}|0><ph|\hat{Y}_{k}]|0>
\end{eqnarray}
where $K_p$ and $K_h$ are projections of the momentum of particle and hole
states onto quantization axis of the system.

The case of responses is more involved in the sense that matrix elements of
the second operator in the commutator are transition densities which are generally
complex. However, the first operator in the commutator still has real
(for T-even $\hat{A}$) or image
(for T-odd $\hat{A}$) matrix elements and so the averages can be finally reduced
to
\begin{eqnarray}\label{eq:AB_T_ev}
<0|[\hat{Q_k},\hat{J}_{\alpha}]|0> &=& 4i  \sum_{ph}^{K_p,K_h>0}
<ph|\hat{Q_k}|0> \Im{<ph|\hat{J}_{\alpha}|0>} ,
\\
<0|[\hat{P_k},\hat{J}_{\alpha}]|0> &=&
-4 \sum_{ph}^{K_p,K_h>0}
<ph|\hat{P_k}|0> \Re{<ph|\hat{J}_{\alpha}|0>} ,
\end{eqnarray}
where $\Im$ and $\Re$ result in image and real parts of the transition densities.

The matrix elements for operator $\hat{P_k}$ read
\begin{equation}
<ph|\hat{P}_k|0> = i2 \varepsilon_{ph} <ph|\hat{Q}_k|0> - <ph|\hat{Y}_k|0> .
\end{equation}

\end{document}